\documentclass[10pt,twocolumn,letterpaper]{article}

\usepackage{cvpr}
\usepackage{times}
\usepackage{epsfig}
\usepackage{graphicx}
\usepackage{amsmath}
\usepackage{amssymb}
\usepackage{booktabs}

\usepackage[breaklinks=false,bookmarks=false]{hyperref}

\begin{document}

\title{Histo-Genomic Knowledge Distillation For Cancer Prognosis From Histopathology Whole Slide Images}

\author{Zhikang Wang\textsuperscript{1}, Yumeng Zhang\textsuperscript{2} , Yingxue Xu\textsuperscript{3}, Seiya Imoto\textsuperscript{4}, Hao Chen\textsuperscript{3}, Jiangning Song\textsuperscript{1}\thanks{Corresponding author.} \\
\textsuperscript{1}Monash University, \textsuperscript{2}Shanghai Jiao Tong University, \textsuperscript{3}HKUST, \textsuperscript{4}University of Tokyo \\
{\tt\small zhikang.wang@monash.edu, zhangyumeng1@sjtu.edu.cn, yxueb@cse.ust.hk, imoto@hgc.jp, }\\ {\tt\small jhc@cse.ust.hk, jiangning.song@monash.edu}
}

\maketitle

\begin{abstract}
Histo-genomic multi-modal methods have recently emerged as a powerful paradigm, demonstrating significant potential for improving cancer prognosis. However, genome sequencing, unlike histopathology imaging, is still not widely accessible in underdeveloped regions, limiting the application of these multi-modal approaches in clinical settings. To address this, we propose a novel Genome-informed Hyper-Attention Network, termed G-HANet, which is capable of effectively distilling the histo-genomic knowledge during training to elevate uni-modal whole slide image (WSI)-based inference for the first time. Compared with traditional knowledge distillation methods (i.e., teacher-student architecture) in other tasks, our end-to-end model is superior in terms of training efficiency and learning cross-modal interactions. Specifically, the network comprises the cross-modal associating branch (CAB) and hyper-attention survival branch (HSB). Through the genomic data reconstruction from WSIs, CAB effectively distills the associations between functional genotypes and morphological phenotypes and offers insights into the gene expression profiles in the feature space. 
Subsequently, HSB leverages the distilled histo-genomic associations as well as the generated morphology-based weights to achieve the hyper-attention modeling of the patients from both histopathology and genomic perspectives to improve cancer prognosis.  Extensive experiments are conducted on five TCGA benchmarking datasets and the results demonstrate that G-HANet significantly outperforms the state-of-the-art WSI-based methods and achieves competitive performance with genome-based and multi-modal methods. G-HANet is expected to be explored as a useful tool by the research community to address the current bottleneck of insufficient histo-genomic data pairing in the context of cancer prognosis and precision oncology. The Pytorch implemented code is available at \textit{https://github.com/ZacharyWang-007/G-HANet}. 
\end{abstract}

\section{Introduction}
Survival analysis delves into the duration until a specific event happens and has been extensively applied in biology, engineering, economics, and medicine \cite{kleinbaum1996survival}. Particularly in the field of cancer research \cite{cox1972regression,collett2023modelling}, experts can estimate survival status, compare survival rates among different patient groups, and identify critical prognostic factors. Given that cancer stands as a leading cause of human mortality worldwide \cite{ferlay2019estimating}, cancer survival prediction has immense clinical value, thereby drawing increasing research attention.

In current clinical practice, histopathology whole slide image (WSI) and genomic data are widely utilized for cancer prognosis. WSIs provide qualitative insights into the morphology and spatial structure of tumor tissues, as well as their interactions with the adjacent microenvironment; whereas genomic data delivers a quantitative molecular overview from the microcosmic perspective. Both two varieties of data have their unique strengths and constraints. In recent years, a large number of WSI-based \cite{li2018graph,chen2021whole,wang2023surformer} and genome-based \cite{chaudhary2018deep,ching2018cox,wang2019cancer,xing2022multi} uni-modal algorithms have been proposed for cancer prognosis. Moreover, many histo-genomic multi-modal algorithms \cite{yao2017deep,chen2021multimodal,li2022hfbsurv,chen2022pan,xu2023multimodal} have demonstrated superior capabilities in precision medicine, creating a more comprehensive and personalized representation that aligns closely with the patient's health status. Nevertheless, genome sequencing still remains costly for most cancer patients, especially in underdeveloped areas. Hence, the broader implementation of multi-modal algorithms in clinical practice and research could be stalled, preventing their potential benefits from being realized on a larger scale and across diverse demographic groups.

Maximizing the utility of existing histo-genomic pairs to enhance WSI-based network performance is a promising strategy to address the discussed data bottleneck. However, this issue remains unexplored in current computational pathology. Given the success of knowledge distillation (KD) \cite{hinton2015distilling,huang2022knowledge,kim2021comparing,heo2019knowledge,komodakis2017paying,passalis2018learning} in many tasks, a potential solution to address the above issue is to transfer the knowledge from multi-modal models to uni-modal ones. Specifically, we need a powerful multi-modal network as a teacher and a WSI-based uni-modal network as the student. 
However, most response-based knowledge distillation (KD) methods \cite{hinton2015distilling,huang2022knowledge,kim2021comparing} are primarily designed for conventional classification tasks and may not be effective in survival prognosis due to the distinct optimization target. In terms of representation-based methods \cite{heo2019knowledge,komodakis2017paying,passalis2018learning}, aligning the feature distributions between the teacher and student networks may fall short in comprehensively transferring the multi-modal knowledge in terms of interactions and functions. Additionally, training two networks, i.e., multi-modal and uni-modal networks, can be laborious and computationally intensive. Consequently, the current KD strategies designed for most computer vision tasks are not the optimal solution to this issue.

To address the challenge highlighted, we propose the genome-informed hyper-attention network (G-HANet). G-HANet notably elevates the performance in uni-modal WSI-based inference by distilling histo-genomic knowledge from the paired data in the training phase. Specifically, the network comprises the cross-modal associating branch (CAB) and hyper-attention survival branch (HSB). Through the functional genomic data reconstruction from WSIs, CAB can effectively distill the associations between functional genotypes and morphological phenotypes and provide insights into the gene expression profiles in the feature space. To reveal this association, the multi-head cross-attention (MHCA) \cite{vaswani2017attention} module is incorporated into the reconstruction phase to offer the pairwise histo-genomic score. Recognizing the complementary characteristics of genome data and histopathology slides during survival analysis, HSB subsequently utilizes both the derived histo-genomic associations and generated morphology-based weights to achieve the hyper-attention modeling of the patients using WSIs from both histopathology and genomic perspectives. The extracted features from HSB will be utilized for the final survival prognosis. Given that genomic data is exclusively utilized to optimize CAB during training, merely WSIs are required for model inference, hence, our approach offers an efficient solution to the challenge of lacking genomic data.

To evaluate the proposed G-HANet, extensive experiments are conducted on five benchmarking TCGA datasets. Our proposed G-HANet significantly outperforms current WSI-based methods and achieves competitive performance with other genome-based and multi-modal methods. Moreover, our study for the first time considers the unavailability of genome data for cancer prognosis in clinical practice and aims to enhance the WSI-based prognosis by leveraging knowledge distilled from training data pairs, thereby holding substantial importance for the computational pathology community.

\begin{figure*}[t]
  \centering
   \includegraphics[width=0.95\linewidth]{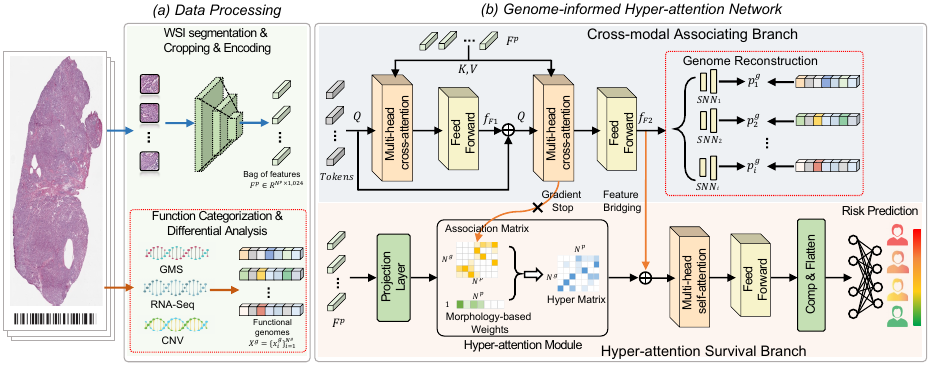}
   \caption{Overview of the proposed genome-informed hyper-attention network (G-HANet). The WSIs and genomic data are formulated as bags of features $F^p$ and categorized according to the functions as $X^g$. In G-HANet, the Cross-modal Associating Branch distills the histo-genomic knowledge through genome reconstruction from WSIs. Afterward, with the assistance of the distilled knowledge, HSB achieves the hyper-attention modeling of the WSIs from both histopathology and genomic perspectives and conducts survival prediction. The genome data processing and reconstruction (framed by the red boxes) are only required in the training phase, thereby enabling the WSI-based inference. 
   }
   \label{Fig1}
\end{figure*}

\section{Related Work}
\subsection{Uni-modal for Cancer Survival Prognosis}
\textbf{Genome-based Methods.} 
Genomic data quantifies the molecular profiles of patients from the microcosmic perspective in an explicit manner. Xu et al. \cite{xu2012gene} utilized the support vector machine-based recursive feature elimination (SVM-RFE) approach to discover the critical gene signatures and conduct prognosis prediction for breast cancer. Chaudhary et al. \cite{chaudhary2018deep} specifically investigated the deep learning-based model to differentiate survival subpopulations of liver cancer patients in six cohorts. This is the first deep-learning study to identify multi-omics features linked to the differential survival of patients with hepatocellular carcinoma. Ching et al. \cite{ching2018cox} developed the Cox-nnet to predict patient prognosis from transcriptomics data and found functional biological insights. Wang et al. \cite{wang2019cancer} and Xing et al. \cite{xing2022multi} employed the graph neural network to model the genomic data for cancer survival prediction based on sample similarity and co-expression gene matrix, respectively.

\textbf{WSI-based Methods.}
WSIs qualitatively offer insights into the tumor characteristics and its adjacent microenvironment.  
Given their gigapixel resolutions, existing methods \cite{li2018graph,chen2021whole,lu2021data,shao2021transmil,zhang2022dtfd,wang2023surformer} generally model them through the weakly supervised multiple instance learning (MIL) strategy \cite{maron1997framework}. Li et al. \cite{li2018graph} explored the significant role of topological features of WSIs in survival analysis by modeling each WSI as a graph. Afterward, the graph convolutional neural network with attention learning was adopted to proceed with the survival prediction. Chen et al. \cite{chen2021whole} proposed the context-aware and spatially-resolved patch-based graph convolutional network, termed Patch-GCN, to hierarchically aggregate instance-level histology features in the tumor microenvironment. Wang et al. \cite{wang2023surformer} proposed a Surformer that can quantify specific histological patterns through bag-level labels without any patch/cell-level auxiliary information. 

Both WSIs and genomic data have their unique strengths and can be synergistically complementary to each other. Merely relying on one data modality will inevitably lose some critical information of patients, leading to suboptimal prognosis.

\subsection{Multi-modal for Cancer Prognosis}
Recent histo-genomic multi-modal methods have demonstrated significant success in the field of precision medicine \cite{yao2017deep,wang2021gpdbn,chen2020pathomic,chen2021multimodal,jaume2023modeling,li2022hfbsurv,chen2022pan,lipkova2022artificial,zhou2023cross}. Yao et al. \cite{yao2017deep} proposed the Deep Correlational Survival Model (DeepCorrSurv), which comprises view-specific and common sub-networks, for integrating multi-view data. They explicitly maximize the correlation among the views and transfer the learned knowledge for the survival regression task. Chen et al. \cite{chen2021multimodal} proposed the Multimodal Co-Attention Transformer (MCAT) framework to learn an interpretable, dense co-attention mapping between WSIs and genomic data. Li et al. \cite{li2022hfbsurv} proposed a novel hierarchical multimodal fusion approach termed HFBSurv. HFBSurv employs a factorized bilinear model to fuse genomic and image features and connect multi-modal information from low-level to high-level. Chen et al. \cite{chen2022pan} utilized multimodal deep learning to jointly examine WSIs and molecular profile data from 14 cancer types and discovered prognostic features correlated to the outcomes. Lipkova et al. \cite{lipkova2022artificial} specifically discussed three different fusion strategies in multi-modal learning and emphasized the enormous potential of personalized precision medicine. 


Despite the promising performance, the clinical application of these methods can be hindered by the shortage and associated sequencing costs and examinations. Therefore, in this study, we aim to leverage the available histo-genomic data pairs to develop a new and powerful deep network with enhanced performance to facilitate WSI-based inference and improve cancer prognosis.

\section{Method}
\subsection{Overview and Problem Formulation}
As illustrated in Fig. \ref{Fig1}, the framework comprises the data processing step and G-HANet survival prognosis. Initially, we convert the WSIs into bags of histopathology features and categorize gene profiles according to their functions. Afterward, the processed bags are fed into the G-HANet for functional gene reconstruction and survival prediction. In this process, the cross-modal associating branch distills the histo-genomic associations and generates insightful features reflecting gene expression profiles. Subsequently, the hyper-attention survival branch utilizes the distilled associations and generated morphology-based weights to model the patients using WSIs from both histopathology and genomic perspectives. It is worth noting that genome-correlated operations in the red boxes (i.e., genome sequencing, function categorization, and genome reconstruction) will be discarded during the inference stage, which enables the WSI-based uni-modal inference.

\textbf{WSI bags formulation.}
WSIs inherently have gigapixel resolutions, which pose considerable challenges for end-to-end processing using current hardware. Building on previous studies \cite{wang2023targeting,chen2021multimodal,chen2022pan}, we employ the weakly supervised multiple-instance learning (MIL) strategy \cite{maron1997framework}, through which each WSI can be formulated as multiple instances assigned with one bag-level label. We initially utilize the automated segmentation algorithm to distinguish the tissue regions from the background and then crop the informative tissues into a series of patches $X^p=\{ x_{i}^p \}_{i=1}^{N_p}$. Eventually, the $N_p$ cropped patches are converted into $d$-dimensional feature vectors through a pre-trained encoder $E$ as:
\begin{equation}
\setlength\abovedisplayskip{3pt}
\setlength\belowdisplayskip{3pt}
    F^p=\{E(x_i^p): x_i^p \in X^p \} = \{f_i\}_{i=1}^{N_p} \in \mathbb{R}^{N_p \times d}.
\end{equation}

\textbf{Genome Pre-processing.} 
Genome profiles, e.g., gene mutation status, copy number variation (CNV), and bulk RNA-Seq abundance, are quantified as 1x1 measurements. As such, they cannot directly convey semantic characteristics in terms of functions or signatures in the context of genome analysis. In this study, we build upon the previous works \cite{chen2021multimodal,xu2023multimodal} and categorize the genes by referring to the six genomic functions: 1) tumor suppression, 2) oncogenesis, 3) protein kinases, 4) cellular differentiation, 5) transcription, and 6) cytokines and growth. In G-HANet, we aim to distill the histo-genomic associations by reconstructing functional genes from WSIs. Given that only a few genes are highly related to the cancer prognosis, reconstructing all the genome data could potentially increase the unnecessary training burden of the neural networks, leading to inferior performance. Consequently, we first conducted the differential analysis using the DESeq2 \cite{love2014moderated} to identify the differential genes between the high-risk and low-risk patients in the training data. Here, the two patient groups were categorized according to their survival status at the middle time point. The detailed sequence length variations across functions and cancer types can be found in Table \ref{table1}. Hereby, the final functional genes can be represented as $X^g = \{x^g_i\}_{i=1}^{N_g}$.

\subsection{Cross-modal Reconstruction for Knowledge Distillation}
The cross-modal associating branch (CAB) is designed to distill the histo-genomic association matrix and generate functional features. Here, the matrix can associate the genotypes and phenotypes, and the functional features can reflect the gene expression profiles in the feature space, both of which will be fed into HSB for survival prognosis. To achieve this target, we reconstruct the functional genes $X^g$ from WSIs, in which way the deep correlations between the two modalities can be automatically established. As illustrated in Fig. \ref{Fig1}, CAB proceeds with feature extraction through two multi-head cross-attention (MHCA) modules with each accompanied by a feed-forward network (FFN), and genome reconstruction through $N_g$ Self-Normalizing Networks (SNNs) \cite{klambauer2017self}.

Regarding the first MHCA, it takes learnable tokens $T$ and the bag of feature $F^p$ as input. In this process, $T$ targets collecting genomic function-related morphological features and will be optimized throughout the training process. The query ($Q$) is derived from $T$, and the key ($K$) and value ($V$) are derived from $F^p$, all of which are achieved through linear transformations. The multi-head mechanism is employed to segment the features into $n$ portions along the feature dimension, generating $Q \in \mathbb{R}^{N_g \times n \times (d'/n)}$ and $K, V \in \mathbb{R}^{N_p \times n \times (d'/n)}$. Subsequently, a cross-product operation between $Q$ and $K$ is executed to derive the attention matrix and a $softmax$ operation is applied to scale the attention values. For this explanation, we assume $n=1$ and formulate this process as follows:
\begin{equation}
\setlength\abovedisplayskip{3pt}
\setlength\belowdisplayskip{3pt}
   m^{1'}_{i,j}=\frac{exp(m_{i,j})}{\sum_{j=1}^{N_p}exp(m_{i,j})}, \quad m_{i,j}^1=\frac{Q_i \times K_j}{\sqrt{d_k}},
\end{equation}
where $\sqrt{d_k}$ is a scaling factor and, $i$ and $j$ present the indices of $Q$ and $K$, respectively. Each element of the matrix, e.g., $m_{i,j}$, serves as an indirect marker of the relative likeness between $Q_i$ (derived from $T$) and $K_j$ (derived from $F^p$) in the feature space, which essentially reflects the pair-wise association between the $i_{th}$ genomic function and $j_{th}$ patch. Afterward, we aggregate the detected features through the cross-product operation between $V$ and $m'$ and transform the features with a linear layer.

Upon aggregating the features, we employ the FFN for the feature nonlinear transformation, getting functional features $f_{F1}$. 
Although the tokens $T$ can be optimized towards different functional genes, collecting critical features from large amounts of patches could be challenging ($N_p \gg N_g$). Here, we employ the summation of $T$ and aggregated features $f_{F1}$ as the query in the second MHCA for another round of detection and aggregation. 
In this way, the query consists of knowledge from both optimized $T$ and the last computation round, thereby ensuring more sensitivity towards correlative features. We regard the attention matrix before the $softmax$ operation as the final association matrix ($M^{am}$). After the second detection round, the features will be fed into another FFN for transformation and produce the final functional features $f_{F2}$, which will be utilized for subsequent genome reconstruction. The whole process can be represented as follows:
\begin{equation}
\setlength\abovedisplayskip{3pt}
\setlength\belowdisplayskip{3pt}
\begin{aligned}
    &f_{F1} = FFN_0(MHCA_0(T, F^p)); \\
    &f_{F2} = FFN_1(MHCA_1(T+f_{F1}, F^p)).
\end{aligned}
\end{equation}
Notably, the two MHCAs share the parameters to facilitate and stabilize the model training.

Given the varying lengths of functional genes, a specific self-normalizing network (SNN \cite{klambauer2017self})-based prediction sub-branch is designated for each category. SNN is a powerful network in modeling high-dimensional low-sample size (HDLSS) scenarios. Here, each SNN sub-branch is constructed by two linear layers, with the first followed by a layer normalization ($LN$) \cite{ba2016layer} and $ELU$ activation function \cite{clevert2015fast}. We denote the reconstructed functional genes as $P^g=\{p_i^g\}_{i=1}^{N_g}$.

\begin{figure}
  \centering
   \includegraphics[width=0.95\linewidth]{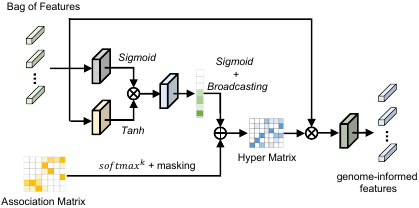}
   \caption{Overview of the hyper-attention module. It takes the transformed bag of features $F^p$ and association matrix $m$ as input and outputs the aggregated genome-informed features $f_{GI}$. 
   }
   \label{Fig2}
\end{figure}

\subsection{Histo-genomic Hyper-attention for Prognosis}
The hyper-attention survival branch (HSB) aims to generate comprehensive prognosis representations for patients through hyper-attention WSI modeling and feature long-range dependency establishment. Specifically, the proposed hyper-attention module (HM) considers the complementary characteristics of different data modalities and aggregates patch features from both histopathology and genomic perspectives. The aggregated features concatenated with $f_{F2}$ will be fed into the multi-head self-attention (MHSA) module and FFN for dependency establishment across features, and into a classification layer for survival prediction. In the following parts, we will introduce the HM and survival prediction process in detail.

\textbf{Hyper-attention Module.}
The proposed hyper-attention module is shown in Fig. \ref{Fig2}. Specifically, it takes both $F^p$ and $M^{am} \in R^{N_p \times N_g}$ (from the second MHCA in CAB) as input. For the attention from the histopathology perspective, we utilize the gated attention network (GAN) \cite{ilse2018attention} to generate the morphology-based weights ($w^{mor}$) for patches. Here, we first feed $F^p$ into two parallel linear layers for feature transformation, each of which is followed by the $sigmoid$ or $tanh$ activation functions, respectively. Afterward, the dot product of the two transformed features is fed into another linear layer to generate the patch-level attention weights. We regard the attention weights after $softmax$ operation along the instance dimension as the final morphology-based weights $ w^{mor} \in \mathbb{R}^{N_p \times 1}$. In terms of the attention from the genomic perspective, we particularly operate $M^{am}$. We apply the $softmax$ operation on the top $k\%$ high-score patches that correspond to each category of gene function and mask the rest as zeros, thereby getting normalized $M^{am'}$ ($\sum_{j=1}^{N_p}M_{i,j}^{am'}=1 $). The gradient of $M^{am'}$ was detached to maintain the learning autonomy of CAB. 
The final hyper matrix $M_{hyper}$ is the mean of $ w^{mor}$ and processed $M^{am}$ and the whole process can be formulated as:
\begin{equation}
\setlength\abovedisplayskip{3pt}
\setlength\belowdisplayskip{3pt}
    M_{hyper} = \frac{w^{mor} + softmax^{k\%}(M^{am})}{2}.
\end{equation}
We take the cross product of the $M_{hyper}$ and $F^p$ after a linear transformation as the genome-informed features $f_{GI}$.

\textbf{Survival Prediction.} 
We concatenate the genome-informed features $f_{GI}$ and $f_{F2}$ as the feature matrix $f_m \in R^{N^g \times d''}$ to represent the WSI. This bridging operation not only enriches the features for survival analysis but also establishes the synergy between the two branches for a joint gradient backpropagation. 
Then, the feature matrix will be fed into the MHSA and FFN for further computation. Compared with MHCA, the query ($Q$), key ($K$), and value ($V$) of MHSA are all derived from the identical input and there is a residual operation between the input and output features. 
Afterward, we employ a linear layer followed by a $LN$ \cite{ba2016layer} and a $ReLU$ activation function for feature dimension compression as $f_m' \in R^{N_g \times d'''}$. 
Rather than employing a readout function (mean or max) for aggregating the first dimension of $f_m'$, we flatten the features into a single features vector. In this way, the features corresponding to different functional genes can be presented in a series connection and the semantic information can be better spatially preserved.

\subsection{Loss Function}
In this study, we aim to proceed with genome reconstruction and survival prediction simultaneously. The genome profiles corresponding to certain functions have very high feature dimensions with some exceeding 1,500, indicating that precisely predicting all the functional genes is extremely challenging. As a result of this, our CAB targets both the gene expression levels and the functional genomic distribution using the mean square error loss $\mathcal{L}_{MSE}$ and scaled cosine error loss $\mathcal{L}_{SCE}$ \cite{hou2022graphmae}, respectively. 
$\mathcal{L}_{SCE}$ \cite{hou2022graphmae} has demonstrated its superiority in reconstructing high-dimensional features in the self-supervised learning setting. Specifically, it can be formulated as follows:
\begin{equation}
\setlength\abovedisplayskip{3pt}
\setlength\belowdisplayskip{3pt}
\mathcal{L}_{SCE} = \frac{1}{|P^g|} \sum_{i=1}^{N^g} (1-\frac{(p_i^g)^T  x_i^g}{||p_i^g||\cdot ||x_i^g||} )^{\gamma },
\end{equation}
where $\gamma>=1$ controls the feature constraints, $x_i^g$ and $p_i^g$ indicates the ground truth and predicted genes. 
Regarding the hazard loss, we follow the previous works \cite{chen2021multimodal} and utilize the negative likelihood loss $\mathcal{L}_{NLL}$ \cite{zadeh2020bias} for optimizing the survival branch. $\mathcal{L}_{NLL}$ can be formulated as follows:
\begin{equation}
\setlength\abovedisplayskip{3pt}
\setlength\belowdisplayskip{3pt}
\begin{aligned}
    \mathcal{L}_{NLL} = & -c_i \cdot log(f_{surv}(y_i, F_i^p)) \\
    & - (1-c_i) \cdot log(f_{surv}(y_i-1, F_j^p)) \\
    &- (1-c_i) \cdot log(f_{hazard}(y_i, F_j^p)), 
\end{aligned}
\end{equation}
where $c$ and $y$ indicate the censorship and overall survival of the patient, respectively. $f_{surv}$ represents the accumulated hazard rates of the patient: 
\begin{equation}
\setlength\abovedisplayskip{3pt}
\setlength\belowdisplayskip{3pt}
    f_{surv}(y_i, F_i^p) = \prod_{j=0}^{y_i}(1-f_{hazard}(F_i^p)[j]). 
\end{equation}

The overall loss function is:
\begin{equation}
\setlength\abovedisplayskip{3pt}
\setlength\belowdisplayskip{3pt}
    \mathcal{L} = \mathcal{L}_{NLL} + \alpha (\mathcal{L}_{MSE} + \mathcal{L}_{SCE}),
\end{equation}
where $\alpha$=0.3 is a hyperparameter to balance the reconstruction and survival prediction.

\begin{table}[ht]
  \centering
  \setlength\tabcolsep{3pt}
  \caption{Details of the five datasets in terms of genomic sequence lengths before and after differential analysis.}
    \scalebox{0.85}{
  \begin{tabular}{lccccc }
    \toprule
    \toprule
    \multicolumn{6}{c}{\underline{Original sequence lengths}} \\
    Function category & BLCA & BRCA & GBMLGG & LUAD & UCEC \\
     \hline
     tumor suppression & 94 & 91 & 84 & 89 & 3 \\
     oncogenesis & 334 & 353 & 314 & 334 & 24 \\
     protein kinases & 521 & 553 & 498 & 534 & 21 \\
     cellular differentiation & 468 & 480 & 415 & 471 & 22 \\
     transcription & 1,495 & 1,565 & 1,395 & 1,509 & 64 \\
     cytokines and growth & 479 & 482 & 428 & 482 & 15 \\
     \hline
     \multicolumn{6}{c}{\underline{Differential analyzed sequence lengths}} \\
     Function category & BLCA & BRCA & GBMLGG & LUAD & UCEC \\
     tumor suppression & 20.2 & 30.2 & 34.6 & 20.0 & 1.2 \\
     oncogenesis & 73.8 & 123.2 & 129.8 & 87.8 & 8.4 \\
     protein kinases & 102.2 & 109.6 & 177.4 & 146.4 & 7.2 \\
     cellular differentiation & 94.8 & 171.4 & 139.8 & 171.4 & 8.0 \\
     transcription & 351.6 & 538.6 & 535.2 & 386.2 & 24.0 \\
     cytokines and growth & 78.2 & 165.8 & 126.4 & 128.2 & 5.4 \\
    \bottomrule
    \bottomrule
  \end{tabular}}
  \label{table1}
\end{table}

\section{Experiments}
\subsection{Datasets and Evaluation Metrics}
Following \cite{chen2021multimodal,xu2023multimodal}, extensive experiments of this study were conducted on five different cancer types from The Cancer Genome Atlas (TCGA) including Bladder Urothelial Carcinoma (BLCA), Breast Invasive Carcinoma (BRCA), Glioblastoma \& Lower Grade Glioma (GBMLGG), Lung Adenocarcinoma (LUAD), and Uterine Corpus Endometrial Carcinoma (UCEC) datasets. 5-fold cross-validation with the 4:1 ratio of training and validation sets is performed on each dataset. The mean concordance index (c-index) with its standard deviation (std) is reported for a comprehensive performance comparison. 


Table \ref{table1} specifically details the sequence lengths of initial and after differential analysis of the five datasets across six functional categories (tumor suppressor genes, oncogenes, protein kinases, cell differentiation markers, transcription, and cytokines and growth). It is important to note that the differential analysis was conducted on the training samples of each fold. As a result, the number of significant genes identified can vary, and only the mean sequence lengths are reported. Owing to the markedly reduced number of genes in the UCEC dataset relative to the other four datasets, differential analysis was not performed while training the UCEC-based model.


\begin{table*}
\vspace{-2.0em}
  \centering
  \normalsize
  \caption{Performance comparison with state-of-the-art genome-based/WSI-based uni-modal algorithms and histo-genomic multi-modal algorithms on TCGA datasets in terms of c-index. The 'Patho.' (WSI) and 'Geno.' (genomic data) indicate the data utilized during inference. 'KD' represents the knowledge distillation strategy. We label the best WSI-based inference model with red color and highlight the best performance in bold format. }
  \scalebox{0.85}{
  \begin{tabular}{l*{9}{c}}
    \toprule
    \toprule
    Methods & Patho. & Geno. & KD & BLCA & BRCA & GBMLGG & LUAD & UCEC & Overall\\
    \hline
    SNN \cite{klambauer2017self} & & \checkmark & &   0.618 ± 0.022  & 0.624 ± 0.060 &0.834 ± 0.012 & 0.611 ± 0.047 &    0.679 ± 0.040 & 0.673\\
    SNNTrans \cite{klambauer2017self,shao2021transmil} & & \checkmark & & 0.659 ± 0.032 & 0.647 ± 0.063 &0.839 ± 0.014 &  0.638 ± 0.022   &  0.656 ± 0.038 & 0.688\\
    \hline
    AttnMIL \cite{ilse2018attention} & \checkmark &&& 0.551 ± 0.049 & 0.577 ± 0.043 &	0.786 ± 0.026 &	0.561 ± 0.078 &	0.639 ± 0.057 & 0.623 \\
    DeepAttnMISL \cite{yao2020whole} & \checkmark &&& 0.504 ± 0.042 & 0.524 ± 0.043 &0.734 ± 0.029 &  0.548 ± 0.050  &  0.597 ± 0.059 & 0.581\\
    CLAM-SB \cite{lu2021data} & \checkmark &&& 0.559 ± 0.034 & 0.573 ± 0.044 & 0.779 ± 0.031 & 0.594 ± 0.063  &  0.644 ± 0.061 & 0.630\\ 
    CLAM-MB \cite{lu2021data} & \checkmark &&& 0.565 ± 0.027 & 0.578 ± 0.032 & 0.776 ± 0.034 & 0.582 ± 0.072  &0.609 ± 0.082 & 0.622\\
    TransMIL \cite{shao2021transmil} & \checkmark &&& 0.608 ± 0.139 & 0.626 ± 0.042 & 0.798 ± 0.033 & \textcolor{red}{0.641 ± 0.033} & 0.657 ± 0.044 & 0.666\\
    DTFD-MIL \cite{zhang2022dtfd} & \checkmark &&& 0.546 ± 0.021 & 0.609 ± 0.059 & 0.792 ± 0.023 & 0.585 ± 0.066 & 0.656 ± 0.045 & 0.638\\
    Surformer \cite{wang2023surformer} & \checkmark &&& 0.594 ± 0.027 & 0.628 ± 0.037 & 0.809 ± 0.026 & 0.591 ± 0.064 & 0.681 ± 0.028 & 0.661\\
    \hline
    Porpoise \cite{chen2020pathomic} & \checkmark & \checkmark && 0.636 ± 0.024 & 0.652 ± 0.042 &0.834 ± 0.017 & 0.647 ± 0.031  &  0.695 ± 0.032 & 0.693\\ 
    MCAT \cite{chen2021multimodal} & \checkmark & \checkmark && 0.672 ± 0.032 & 0.659 ± 0.031 & 0.835 ± 0.024 & 0.659 ± 0.027 & 0.649 ± 0.043 & 0.695\\ 
    MOTCat \cite{xu2023multimodal} & \checkmark & \checkmark && \textbf{0.683 ± 0.026} & \textbf{0.673 ± 0.006} & \textbf{0.849 ± 0.028} & \textbf{0.670 ± 0.038}  &  0.675 ± 0.040 & \textbf{0.710}\\
    \hline
    CLAM-SB \cite{lu2021data,hinton2015distilling} & \checkmark && \checkmark & 0.567 ± 0.026 & 0.601 ± 0.039 & 0.801 ± 0.030 & 0.570 ± 0.021 & 0.668 ± 0.057 & 0.641\\ 
    CLAM-SB \cite{lu2021data,komodakis2017paying} & \checkmark && \checkmark & 0.563 ± 0.066 & 0.589 ± 0.030 & 0.794 ± 0.029 & 0.577 ± 0.095 & 0.640 ± 0.068 & 0.633\\ 
    CLAM-MB \cite{lu2021data,hinton2015distilling} & \checkmark && \checkmark & 0.560 ± 0.047 & 0.585 ± 0.041 & 0.796 ± 0.030 & 0.579 ± 0.053 & 0.608 ± 0.058 & 0.626\\
    CLAM-MB \cite{lu2021data,komodakis2017paying} & \checkmark && \checkmark & 0.618 ± 0.029 & 0.533 ± 0.050 & 0.779 ± 0.026 & 0.578 ± 0.053 & 0.635 ± 0.066 & 0.629\\    
    Ours & \checkmark & & \checkmark & \textcolor{red}{0.630 ± 0.032} & \textcolor{red}{0.664 ± 0.065} & \textcolor{red}{0.817 ± 0.022} & 0.612 ± 0.028 & \textbf{ \textcolor{red}{0.729 ± 0.050}} & \textcolor{red}{0.690}\\
    \bottomrule
    \bottomrule
  \end{tabular}}
  \label{table2}
\end{table*}

    

\begin{table}
\vspace{-1.0em}
  \centering
  \normalsize
  \caption{Analysis $k\%$ in CAB on five TCGA datasets.}
  \scalebox{0.85}{
  \begin{tabular}{l*{7}{c}}
    \toprule
    \toprule
    $k$ & BLCA & BRCA & GBMLGG  \\
    \hline
    10 & 0.619 ± 0.034 & 0.650 ± 0.054 & 0.815 ± 0.021  \\
    15 & 0.625 ± 0.025 & \textbf{0.664 ± 0.065} &  \textbf{0.817 ± 0.022} \\
    20 & \textbf{0.630 ± 0.032} & 0.651 ± 0.063 & 0.814 ± 0.020  \\
    25 & 0.626 ± 0.033 & 0.654 ± 0.064 & 0.811 ± 0.019\\
    30 & 0.624 ± 0.036 & 0.650 ± 0.057 & 0.812 ± 0.021 \\
    35 & 0.620 ± 0.026 & 0.647 ± 0.064 & 0.814 ± 0.022 \\
    \hline
    $k$ & LUAD & UCEC & Overall \\
    \hline
    10 & 0.604 ± 0.041 & 0.706 ± 0.037 & 0.679 \\
    15 & 0.610 ± 0.039 & 0.704 ± 0.030 & 0.684 \\
    20 &  \textbf{0.612 ± 0.028} & 0.717 ± 0.021 & \textbf{0.685} \\
    25 & 0.604 ± 0.045 & 0.721 ± 0.057 & 0.683\\
    30 & 0.612 ± 0.039 & \textbf{0.727 ± 0.056} &  \textbf{0.685}\\
    35 & 0.609 ± 0.041 & 0.729 ± 0.050 & 0.684 \\
    \bottomrule
    \bottomrule
  \end{tabular}}
  \label{table3}
\end{table}

\subsection{Implementation Details}
All the WSIs undergo processing at 20 times (20 $\times$) magnification level. During the pre-processing step, an automated segmentation algorithm was utilized to eliminate the background patches with a low saturation value ($<$ 15). Subsequently, the tissue regions were cropped into a series of non-overlapping patches with a unified 256 $\times$ 256 resolution. Then, a modified ImageNet \cite{deng2009imagenet} pre-trained ResNet50 model \cite{he2016deep} (constructed by one \textit{convolution block} and three \textit{residual blocks}) embedded the original patches into a collection of feature vectors. In the feed-forward computation process, we trained the neural network with the Adam optimizer for 20 epochs in an end-to-end fashion. In each iteration, the batch size was set as 1 and the gradients were accumulated over 32 steps for back-propagation. The learning rate was initialized as 0.0002 for training the BLCA, BRCA, GBMLGG, and LUAD datasets. Regarding the UCEC datasets, the learning rate was set as 0.0001. All experiments were carried out using one NVIDIA GeForce RTX 3090 Card. 

\begin{figure*}
  \centering
   \includegraphics[width=0.95\linewidth]{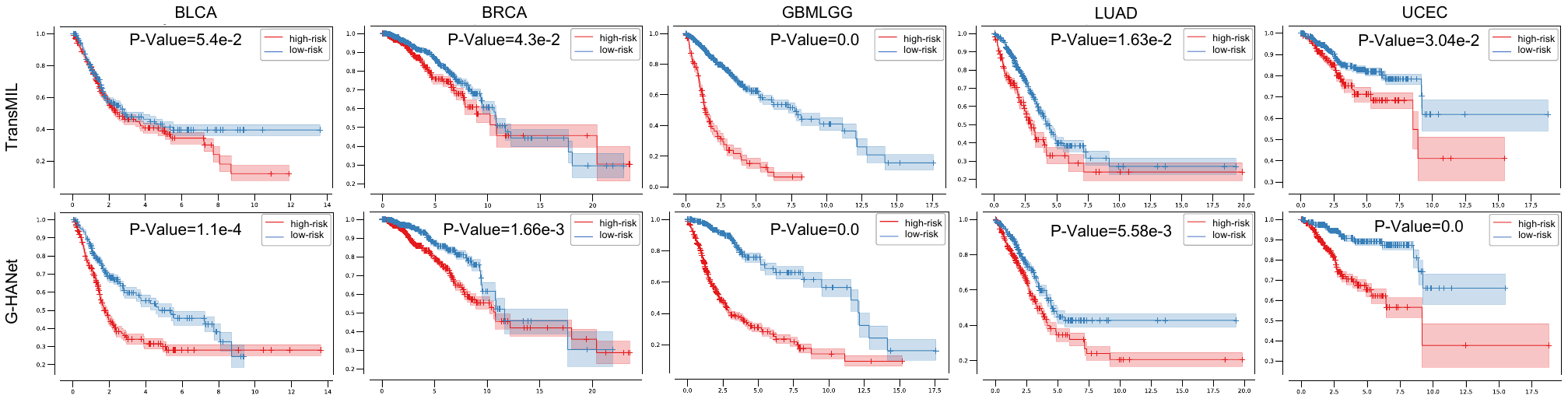}
   \caption{Kaplan-Meier survival curves of TransMIL and our proposed G-HANet on five cancer datasets. High-risk and low-risk patients are indicated with red and blue, respectively. The shaded areas indicate the confidence intervals of each cohort. P-value $<$ 0.05 represents a significant statistical difference between the two groups. 
   }
   \label{Fig3}
\end{figure*}

\begin{table*}
  \centering
  \vspace{-2.0em}
  \normalsize
  \caption{Ablation study of the proposed G-HANet on TCGA datasets in terms of c-index. CAB, CSB, and HSB refer to the cross-modal associating branch, cross-attention survival branch, and hyper-attention survival branch, respectively.
  }
  \scalebox{0.85}{
  \begin{tabular}{l|cc|cc|ccccc}
    \toprule
    \toprule
    Methods & CAB1 & CAB2 & CSB & HSB & BLCA & BRCA & GBMLGG & LUAD & UCEC \\
    \hline

    $Model_0$ &&& &&  0.551 ± 0.049 & 0.577 ± 0.043 & 0.786 ± 0.026 & 0.561 ± 0.078 & 0.639 ± 0.057\\
    \hline
    $Model_1$  & \checkmark  & & \checkmark  &  &  0.587 ± 0.024 & 0.630 ± 0.046 &	0.807 ± 0.026 &	0.584 ± 0.051 &	0.676 ± 0.032\\
    $Model_2$ &&  \checkmark & \checkmark & & 0.614 ± 0.064 & 0.647 ± 0.055 & 0.816 ± 0.027 & 0.612 ± 0.050 & 0.692 ± 0.021 \\
    $Model_3$ & \checkmark  && &\checkmark  & 0.590  ± 0.024 & 0.640  ± 0.051 &  0.808 ± 0.021 & 0.590 ± 0.029 &  0.695 ± 0.030 \\
    \hline
    $Model_4^{*}$ & & \checkmark & & \checkmark  &  0.587 ± 0.048 & 0.646 ± 0.021 & 0.808 ± 0.017 & 0.594 ± 0.038 & 0.698 ± 0.028 \\
    $Model_5^{**}$ & & \checkmark & & \checkmark   &  0.622 ± 0.026 & 0.647 ± 0.050 & 0.816 ± 0.015 & 0.604 ± 0.040 & 0.689 ± 0.045 \\
    $Model_6^{\dag}$ & & \checkmark & & \checkmark   &  0.617 ± 0.048 & 0.643 ± 0.063 & 0.808 ± 0.022 & 0.583 ± 0.041 & - \\
    $Model_7$ & & \checkmark & & \checkmark  &  \textbf{0.630 ± 0.032} & \textbf{0.651 ± 0.063} & \textbf{0.814 ± 0.020} & \textbf{0.612 ± 0.028} & \textbf{0.717 ± 0.021}  \\
    \bottomrule
     \bottomrule
  \end{tabular}}
  \label{table4}
\end{table*}

\subsection{Comparison with State-of-the-art Methods}
In this subsection, we compare our proposed G-HANet with other state-of-the-art uni-modal (WSI-based and genome-based) methods and multi-modal methods in Table \ref{table2}. In terms of genome-based methods, we specifically implement SNN \cite{klambauer2017self} and SNNTrans \cite{klambauer2017self,shao2021transmil}. SNN \cite{klambauer2017self} takes the concatenated genomic profiles as a feature vector and predicts the survival. SNNTrans \cite{klambauer2017self,shao2021transmil} initially categorizes the genomic data according to the functions and then utilizes TransMIL \cite{shao2021transmil} to predict the overall survival. 
As for WSI-based methods, DeepAttnMISL \cite{yao2020whole}, CLAM-SB \cite{lu2021data}, CLAM-MB \cite{lu2021data} are all based on the attention mechanism for representation learning. TransMIL \cite{shao2021transmil} and Surformer \cite{wang2023surformer} are two Transformer-based methods with distinct learning strategies. TransMIL \cite{shao2021transmil} explores the long-range dependencies among patches; whereas Surformer \cite{wang2023surformer} utilizes the prototypes for pattern-specific feature aggregation. DTFD-MIL \cite{zhang2022dtfd} introduces the pseudo bags and employs knowledge distillation for empowering the feature learning capability of the neural networks. 
Regarding multi-modal methods, the attention-based Porpoise \cite{chen2020pathomic}, Transformer-based MCAT \cite{chen2021multimodal}, and optimal-transport-based MOTCat \cite{xu2023multimodal} are introduced for a more challenging comparison. We also implement KD-based \cite{hinton2015distilling,komodakis2017paying} CLAM-SB \cite{lu2021data} and CLAM-MB \cite{lu2021data} with MCAT \cite{chen2021multimodal} as the teacher network.

\textbf{G-HANet vs genome-based methods.}
When comparing G-HANet with SNN \cite{klambauer2017self}, G-HANet surpasses SNN in performance on the BLCA, BRCA, LUAD, and UCEC datasets by 1.2\%, 4.0\%, 0.1\%, and 5.0\%, respectively. Against SNNTrans \cite{klambauer2017self,shao2021transmil}, G-HANet achieves better performance on the BRCA and UCEC dataset. Based on this comparison, we can conclude that with the assistance of genomic data during training, G-HANet can derive critical features from WSIs and achieve competitive and even superior performance than genome-based methods.

\textbf{G-HANet vs WSI-based methods.}
In this comparison, we witness a significant performance improvement of G-HANet compared with other WSI-based methods. On the BLCA, BRCA, GBMLGG, and UCEC datasets, G-HANet outperforms the second-best methods by 2.2\%, 3.6\%, 0.8\%, and 4.8\%, respectively. On the LUAD dataset, TransMIL \cite{shao2021transmil} achieves a c-index of 0.641, surpassing G-HANet in performance by 2.9\%. However, TransMIL \cite{shao2021transmil} has been trapped for its substantial increase in computational demands when applied to large-scale Whole Slide Images (WSIs). Meanwhile, the overall performance of TransMIL \cite{shao2021transmil} on the five datasets is 2.4\% lower than G-HANet. We also implement response-based \cite{hinton2015distilling} and representation-based \cite{komodakis2017paying} KD on CLAM-MB \cite{lu2021data} and CLAM-SB \cite{lu2021data} networks with MCAT \cite{chen2021multimodal} as the teacher. By comparing KD-based methods with other WSI-based methods, it is evident that KD \cite{hinton2015distilling,komodakis2017paying} does not yield benefits across every dataset, and the overall improvement it offers is limited. Furthermore, the representation-based KD \cite{komodakis2017paying} is designed to transfer the knowledge of a big model to a small one with a similar network architecture, therefore, it is not suitable for transferring knowledge from a multi-modal network to a uni-modal one. Overall, we can conclude that our proposed G-HANet is superior to current WSI-based methods and our distilling strategy is more suitable for distilling multi-modal knowledge to elevate uni-modal inference.

\textbf{G-HANet vs multi-modal methods. }
In the comparison of G-HANet with multi-modal methods, it achieves competitive performance on the BRCA dataset and the best performance on the UCEC dataset. On the UCEC datasets, G-HANet surpasses the MOTCat \cite{xu2023multimodal} and Porpoise \cite{chen2020pathomic} by 5.4\% and 3.4\%, respectively. 
We attribute the moderate performance of the multi-modal methods to the limited number of detected genes. As shown in Table \ref{table1} last column, the genes corresponding to each function are mostly below 20. Moreover, not all the genes are correlative to cancer prognosis. Merely relying on the communication between the two modalities to model the patients will lose many critical features, thereby resulting in inferior model performance. Consequently, it is evidenced that G-HANet exhibits strong performance in comparison with the multi-modal methods and demonstrates resilience to variations in genomic lengths.


\begin{figure*}
  \centering
  \vspace{-2.0em}
   \includegraphics[width=0.95\linewidth]{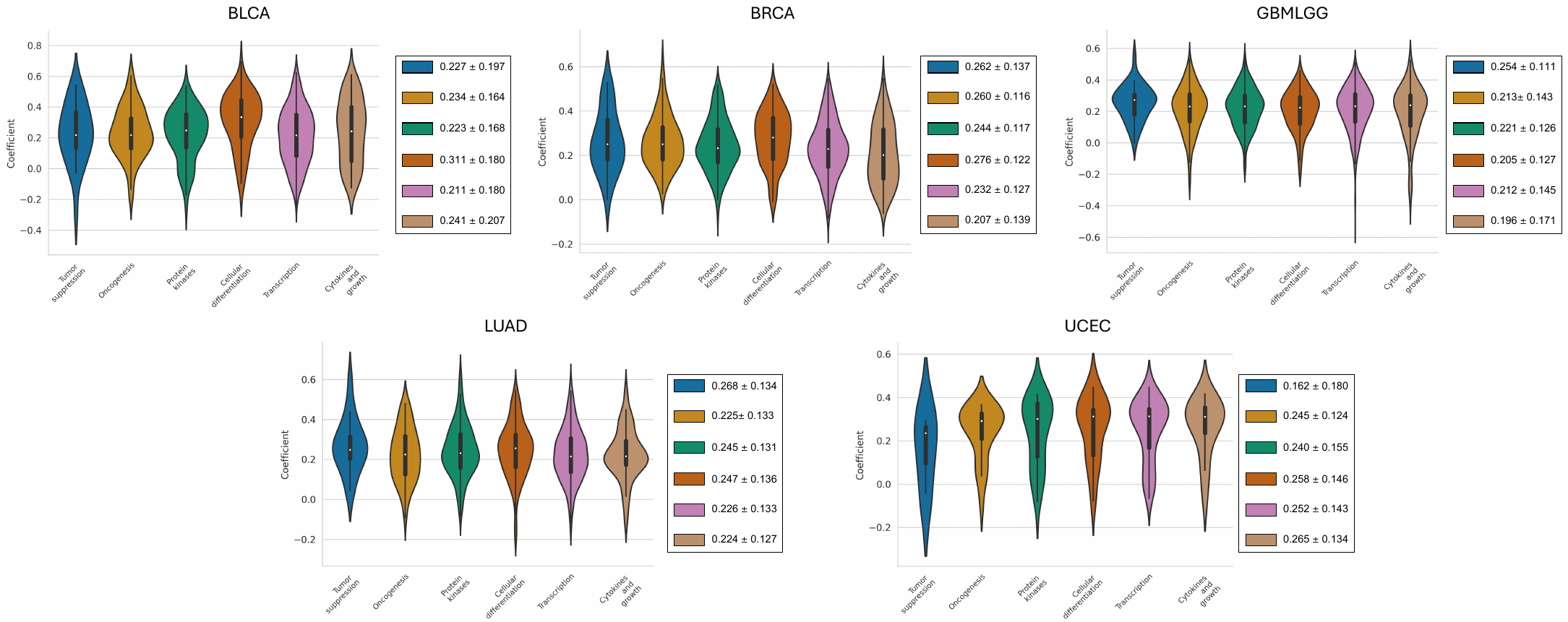}
   \caption{Violin plots of the Spearman Correlation coefficients of six genomic functions on five TCGA datasets. The mean values and corresponding standard variations are placed next to each plot. }
   \label{Fig4}
\end{figure*}

\subsection{Ablation Study}
\textbf{Analysis of top $k\%$ in CAB:} 
In CAB, we select the top $k\%$ patches for generating the genome-informed features. Here, we specifically discuss the impact of $k$ across the five TCGA datasets. As shown in Table \ref{table3}, for different cancer types, the model performs best with different $k$ values.  On BLCA and LUAD datasets, G-HANet achieves the best performance when $k=20$. On BRCA and GBMLGG datasets, the optimal threshold is $k=15$. Regarding the UCEC dataset, the optimal $k$ is 30. When $k=20$ and $30$, we can achieve the best overall performance on the five datasets reaching 0.685. Given the cross-tumor heterogeneity, which leads to variations in characteristics of progression, cell types, and immune responses, it is reasonable to expect that different datasets may have distinct optimal values for $k$.

\textbf{Analysis of the proposed CAB and HSB:} 
In this subsection, we discuss the effectiveness of the proposed CAB and HSB. We set $k=20$ for a fair comparison. The experiment results are shown in Table \ref{table4}. 
$Model_0$ is the baseline that utilizes a gated attention network (GAN) \cite{ilse2018attention} to model the WSIs from the histopathology view. 
Regarding the CAB, our discussion focuses on two main branches: the self-derived-based and the cross-attention-based. Specifically, CAB1 and CAB2 generate the features for the six functional categories through GAN \cite{ilse2018attention} and MHCA (Section III. B), respectively. As for the survival branch, we replace the $M_{hyper}$ in HSB with the MHCA using $f_{F2}$ as the query and term the modified one as the cross-attention survival branch (CSB). Moreover, we indicate the HSB without the feature bridging operation and merely relying on $M^{am}$ for patch aggregation as $Model_4^{*}$ and $Model_5^{**}$, respectively. $Model_6^{\dag}$ did not implement differential analysis on the genomic data. $Model_7$ represents our proposed G-HANet.

Here, we take the model performance on the BLCA as an example. By comparing $Model_0$ with others, we can witness a significant performance gap, which shows that introducing genome data into model training is quite effective. By comparing $Model_1$, $Model_2$, $Model_3$, and $Model_7$, we can find that our CAB2 is much more powerful than CAB1 with $Model_2$ outperforming $Model_1$ by 1.7\% and $Model_7$ outperforming $Model_3$ by 2.0\%. 
By comparing $Model_7$ with $Model_2$, there is also a 2.1\% improvement, indicating that utilizing the $M^{am}$ and $w^{mor}$ jointly for aggregating patches can emphasize more survival-related features than the cross-attention-based method. The comparison of $Model_4^{*}$,  $Model_5^{**}$, and $Model_7$ highlights the effectiveness of feature bridging operation and $M^{am}$ in generating comprehensive representations for patients. Additionally, without genomic data differential analysis, the trained models would have performance decrease to some extent. In all, each proposed module is effective in the representation learning process and can work independently and cooperatively.

\subsection{Genome Prediction Analysis}
In the proposed G-HANet, we explored the histo-genomic associations by genome reconstruction using CAB. Given the fact that predicting the expression values of all the genes (more than 500 genes after differential analysis) from the WSIs is technically quite challenging. Here, we evaluated the prediction performance using the Spearman Correlation, which is a non-parametric measure of the strength and direction of association that exists between two variables measured on an ordinal scale. The coefficient ranges from -1 to +1, where -1, 0, and +1 indicate perfect negative, no, and perfect positive associations of the ranks, respectively. 
As shown in Figure \ref{Fig4}, the mean SC coefficients of the six functional categories on the five datasets generally range from 0.2 to 0.3, which suggests a weak positive linear relationship between the ranks of the two variables being compared. Although the trained CAB cannot accurately predict the expression values for all the genes, the learned hidden correlations could truly help understand the potential functions of patches, thereby improving the characterization of patients from WSIs and enhancing the cancer prognosis accuracy. 

\subsection{Kaplan-Meier Survival Curve} 
The Kaplan-Meier (KM) survival curve serves as a non-parametric approach to display the survival probability over time for one or more groups. In this subsection, we employ it to visualize the differences in survival status between the predicted high-risk and low-risk cohorts. Specifically, the median of the predicted survival rate at the middle time point was utilized as a criterion to categorize the patients into the high-risk and low-risk groups with an equal number of patients. Moreover, the log-rank test is conducted to evaluate the statistically significant difference between the two groups of patients. Specifically, we visualize the KM curves of both TransMIL \cite{shao2021transmil} and our proposed G-HANet on the five datasets in Fig. \ref{Fig3}. 

In all cases, G-HANet exhibits P-values smaller than 5e-2, whereas the P-value for TransMIL on the BLCA dataset exceeds 5e-2. This means that TransMIL trained on the BLCA dataset could not properly estimate the survival status of patients, having bad generalization capability. G-HANet outperforms TransMIL in terms of achieving smaller P-values on four datasets, except for an identical result on the GBMLGG dataset. 
Moreover, G-HANet could better distinguish the two patient groups on the five datasets qualitatively. For example, patients from the BLCA and BRCA datasets could be better identified by G-HANet before 8 and 10 months, respectively. Although TransMIL \cite{shao2021transmil} outperforms G-HANet on the LUAD dataset in terms of c-index, the KM-curve indicates that patients living over 10 months are not significantly distinguishable. Based on the above observations, it is evident that G-HANet excels in distinguishing high-risk and low-risk cohorts and it can emerge as a powerful model for cancer prognosis. 

\begin{figure*}
  \centering
  \vspace{-2.0em}
   \includegraphics[width=0.9\linewidth]{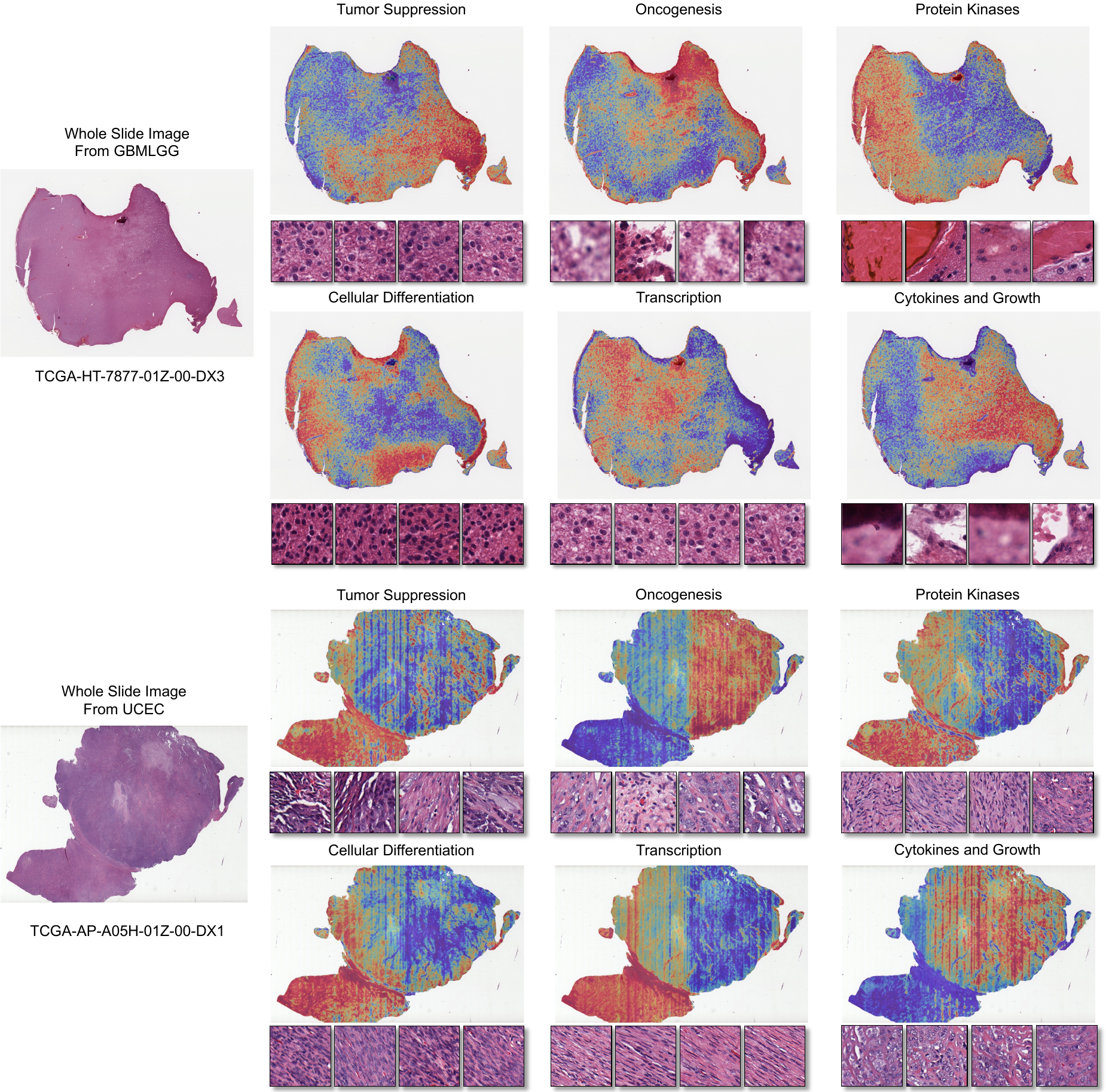}
   \caption{Histo-genomic association visualization from the G-HANet. Here, we particularly visualized the histo-genomic associations of the six different functional genes and presented the top four highlighted patches. Two WSIs from the GBMLGG and UCEC datasets were taken as examples. }
   \label{Fig5}
\end{figure*}

\subsection{Histo-genomic Association Visualization} 
Visualizing the histo-genomic associations derived from the CAB is crucial for model interpretation. In Figure \ref{Fig5}, we employed two WSIs from the UCEC and GBMLGG datasets to depict how specific histopathological patches relate to various functional gene categories, i.e., tumor suppressors, oncogenes, protein kinases, cellular differentiation, transcription, and cytokines. Highlighted top four patches were also retrieved from the original WSI. 

In essence, tumor suppressor genes serve to regulate cell growth and division, whereas oncogenes are mutated or overexpressed versions of normal genes. The two varieties of genes can be seen as opposing in the context of cancer development and progression \cite{kontomanolis2020role}. The illustration provided demonstrates a clear distinction between the morphological patterns associated with tumor suppressor genes and oncogenesis, with minimal overlap observed. Additionally, it was noted that genes related to oncogenesis and cytokines tend to express in similar regions. This observation is consistent with findings that oncogene mutations can have a direct impact on cytokine gene expression \cite{demetri1990expression}. In all, this visualization could aid pathologists in comprehending the model's predictive mechanism, thereby enhancing the credibility and deployment of trustworthy AI systems.

\section{Conclusion}
In this study, we have proposed a novel multi-modal deep neural network, termed Genome-informed Hyper-Attention Network (G-HANet), to improve cancer prognosis. It significantly elevates the network performance on WSI-based inference through multi-modal knowledge distillation of the histo-genomic associations in the training phase. Extensive experiments are conducted on five different TCGA benchmarking datasets and illustrate that G-HANet outperforms current state-of-the-art WSI-based methods and is on par with many genome-based and multi-modal methods. Our approach is tailored to address the current bottleneck of acquiring histo-genomic data pairing during model development and deployment. Future work includes the exploration of a more adaptable algorithm that can accept both uni-modal and multi-modal data to achieve an accurate prognosis.

{\small
\bibliographystyle{ieee_fullname}
\bibliography{egbib}
}

\end{document}